\documentclass[journal=jctcce, manuscript=article]{achemso}

\author{Ahmed A. Maarouf}
\affiliation{IBM T. J. Watson Research Center, Yorktown Heights, NY 10598}
\alsoaffiliation
{Egypt Nanotechnology Research Center, Cairo University, Giza, Egypt}
\email{amaarouf@us.ibm.com}
\author{Razvan A. Nistor}
\affiliation{IBM T. J. Watson Research Center, Yorktown Heights, NY 10598}
\author{Ali Afzali-Ardakani}
\affiliation{IBM T. J. Watson Research Center, Yorktown Heights, NY 10598}
\author{Marcelo A. Kuroda} 
\affiliation{IBM T. J. Watson Research Center, Yorktown Heights, NY 10598}
\author{Dennis M. Newns}
\affiliation{IBM T. J. Watson Research Center, Yorktown Heights, NY 10598}
\author{Glenn J. Martyna}
\affiliation{IBM T. J. Watson Research Center, Yorktown Heights, NY 10598} 
 
\title{Crown Graphene Nanomeshes: Highly Stable Chelation-Doped Semiconducting Materials
}
\keywords{graphene, nanomesh, antidot lattices, doping, ion-chelation, first-principles calculations}

\begin{document}

\begin{abstract} 
Graphene nanomeshes (GNM's) formed by the creation of pore superlattices in graphene, are a possible route to graphene-based electronics due to their semiconducting properties, including the emergence of fractional eV band gaps. The utility of GNM's would be markedly increased if a scheme to stably and controllably dope them was developed. In this work, a chemically-motivated approach to GNM doping based on selective pore-perimeter passivation and subsequent ion chelation is proposed. It is shown by first-principles calculations that ion chelation leads to stable doping of the passivated GNM's -- both {\it n}- and {\it p}-doping are achieved within a rigid-band picture. Such chelated or ``crown'' GNM structures are stable, high mobility semiconducting materials possessing intrinsic doping-concentration control; these can serve as building blocks for edge-free graphene nanoelectronics including GNM-based complementary metal oxide semiconductor (CMOS)-type logic switches.
\end{abstract} 

\section{Introduction}
In the past decade, graphene has been the focus of extensive theoretical and experimental research.\cite{neto2009,geim2007} This one atom layer thick material has unique physical and chemical properties, including high mobility and mechanical strength, that makes it a potential candidate for applications across science and technology, such as nanoelectronics,\cite{graphenetrans1,graphenetrans2,graphenetrans3} RF amplifiers,\cite{graphenetrans4} chemical separation,\cite{Jiang2009} and supports for catalysis.\cite{graphenecatalysis}

Despite the success of graphene in many domains, the material's zero band-gap property has limitations for applications which require a fraction of an eV bandgap and/or a stable doping mechanism, such as computer logic switches. Bilayer graphene in an applied E-field does possess a gap (provided the bilayer can be fabricated in Bernal Stacking)\cite{Xia2010} but the gap is insufficient for room temperature computer switches, for instance. Nanoribbons (long nanometer-thin strips of graphene material), have sufficient band gaps, but in addition to edge scattering, a satisfactory controlled stable doping mechanism has yet to be discovered. Chemical and physical properties of graphene structures with single and double vacancies have also been studied.\cite{Malola2009, Sanyal2009, Topsakal2008, Pisani2008}  

Graphene nanomeshes (GNM's, also called graphene antidot lattices) are novel structures, created by forming a superlattice of pores in a graphene sheet. A GNM is characterized by the details of the pore superlattice (size, superlattice constant, and symmetry of pores), which is superimposed on the intrinsic graphene lattice. These details determine whether the GNM is semimetallic or semiconducting (i.e. the size of the band gap).\cite{Furst2009, Pedersen2008,Liu2009,Petersen2009,Vanevic2009,Martinazzo2010,Baskin2011} The semiconducting GNM can potentially be patterned into devices by spatial control of the pore perimeter passivating species, as we shall discuss.

Several groups have fabricated GNM's and examined their electronic properties.\cite{Bai2010,Safron2011,Liu2011,Liang2010} A block copolymer lithography approach was used to generate GNM's with pore size of about 20-40 nm and with a pore-to-pore width as low as 5 nm.\cite{Bai2010} I-V characteristics of GNM {\it p}-doped field effect transistors (pFET's) showed that their ON-OFF ratio was an order of magnitude higher than that of pristine graphene, although with a lower electrical conductivity.

In order to fabricate CMOS type devices from GNM's it is essential to both {\it n}- and {\it p}-dope the material most simply without major alteration of the band structure. Substitutional doping by ion implantation, which has been demonstrated on graphene,\cite{graphenedoping1,graphenenitrogendoping1,graphenenitrogendoping2,grapheneborondoping1} suffers from concentration fluctuations in nanoscale devices. An alternative strategy realized in graphene is surface doping, wherein non-volatile compounds are spread across a 2D material, but this approach is also subject to undesirable concentration fluctuations and the free moieties (dopants) can migrate across the device leading to stability problems.\cite{graphenedoping1,kim2010} It is therefore desirable to find an approach for doping GNM's that offers both high doping concentration-control and thermal stability.

\section{Chelation Doping}

\begin{figure}
   \includegraphics[width=3.5in]{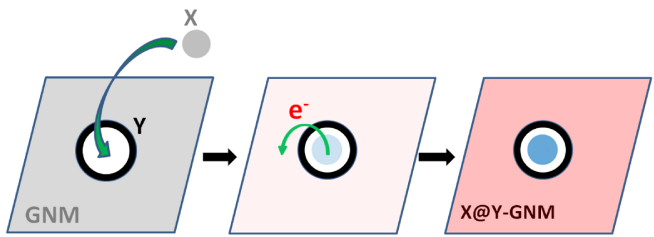}
   \caption
{ 
 A three-state conceptual thermodynamic sketch of the ion chelation doping of a GNM; the chelant atom is brought close to the undoped passivated GNM, resulting in the docking of the chelant in the pore, the ionization of the chelant, and the doping of the GNM, followed by structural relaxation or solvation. Here, {\it n}-doping is shown; {\it p}-doping simply replaces the electron with a hole. The dynamical process involves partial charge transfer and raising/lowering of the HOMO/LUMO of the chelant.\cite{Nistor2011,newns1,newns2}
}
   \label{figure1}
\end{figure}

Here, we utilize a concept from chemistry, ion chelation,\cite{hua2010,indianamolecule} to generate the desired doping physics in GNM's. The carbon atoms at the pore perimeter are chemically active and, unless the GNM is {\it in vacuo}, the pore perimeter will be passivated by a chemical moiety. Ions can be chelated in GNM's because an appropriate passivating species (such as H or O) forms a natural crown in which ions can selectively set or dock like a ``jewel'' (as in the well-known crown-ether materials.\cite{crownethers1,crownethers2}) Passivation with chosen chemical functional groups leads to charge polarization at the pore perimeters due to an electronegativity mismatch between the passivating entity and carbon, thus forming an alluring trap for ions. When an atom approaches a chelation site, the electrostatic energy gain becomes higher than the energy cost of charge transfer, allowing an electron or a hole to transfer to the GNM within a rigid band picture; the resulting ion is tightly bound in the pore. In Fig.~\ref{figure1}, atom X docks in the chelation site which has been passivated by species Y. Charge transfer is stabilized by the electrostatic interaction between ionized species X and the local dipole moment (permanent plus induced) of the pore-perimeter carbon-passivating moiety Y. We term these compounds ``crown GNM's" and give them the symbol X@Y-GNM. 

In order to understand X@Y-GNM physics, we study two GNM passivations, one with electronegativity lower than that of carbon ($\chi_C=2.55$), and another with electronegativity greater than that of carbon. The two natural first choices are hydrogen ($\chi_H=2.0$) and oxygen ($\chi_O=3.44$). Pores passivated with hydrogen form pockets for electron withdrawing {\it p}-dopants, such as fluorine and chlorine. On the other hand, pores passivated with oxygen form pockets for electron donating {\it n}-dopants, such as sodium and potassium.

In detail, we use the local density approximation (LDA) within density functional theory (DFT) to investigate the ion chelation doping of crown GNM's (see supplementary material for computational details, and for results using GGA and hybrid functional based calculations confirming the LDA results). We present results of our electronic structure calculations on four exemplar cases, two for each doping flavor; the {\it p}-doped X@H-GNM through fluorine and chlorine chelation (X=F, Cl), and the {\it n}-doped X@O-GNM through potassium and sodium chelation (X=K, Na). We compare the doped crown-GNM band structure and density of states (DOS) to those of the undoped crown-GNM and graphene, and report the binding energy of each dopant to its chelating GNM.

\begin{figure}
   \includegraphics[width=3.5in]{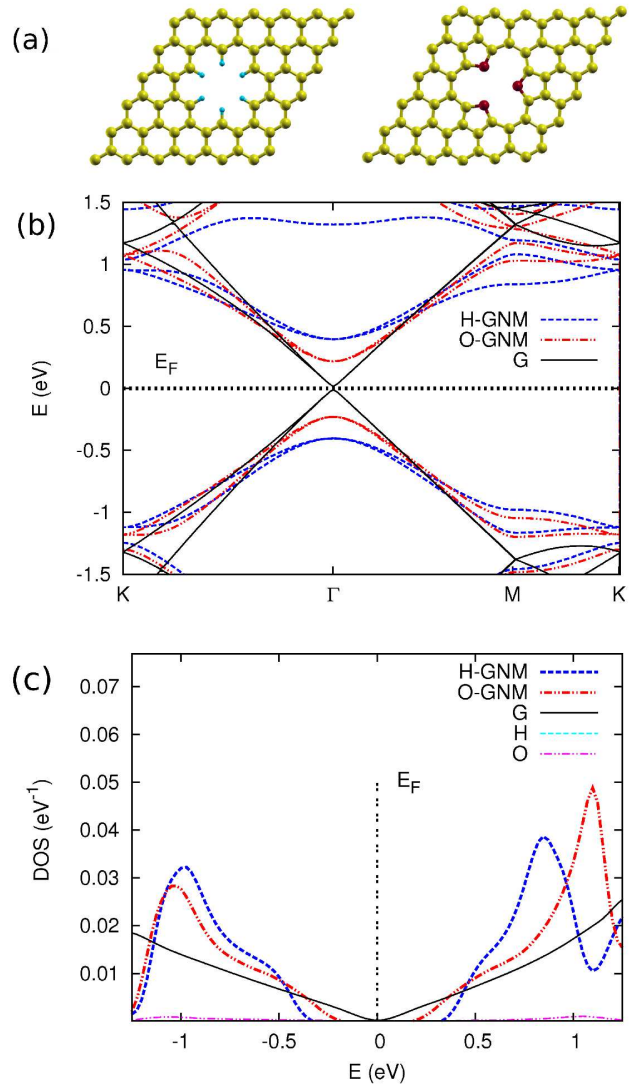}
   \caption
{ 
{\bf (a)} Unit cells of a hydrogen passivated GNM (H-GNM), and an oxygen-passivated GNM (O-GNM), with a triangular lattice of pores and a pore size of about $0.6$ nm. {\bf (b)} band structures of H-GNM (blue, dashed), O-GNM (red, dashed dotted), and pristine graphene (black, solid). {\bf (c)} DOS of: H-GNM (blue, dashed), O-GNM (red, dashed-dotted), and graphene (black, thin-solid). The hydrogen and oxygen contributions to the DOS of the H-GNM and O-GNM, respectively, are also plotted (both have negligible contributions in the energy range shown).
}
   \label{figure2}
\end{figure}

We begin by describing the electronic properties of passivated GNM's, which serve as the foundation for our chelated material. Unpassivated semiconducting GNM's have midgap states resulting from the mixing of the $\sigma$ states of the under-coordinated carbon atoms at the pore perimeter\cite{Furst2009} which are shifted out of the gap by passivation (see supplementary material). In Fig.~\ref{figure2}a, we show a hydrogen- and an oxygen-passivated GNM. We notice that the passivation saturates all atomic bonds, and therefore a passivated GNM is chemically stable (inactive). In Fig.~\ref{figure2}b, we compare the band structures of the two passivated systems and pristine graphene; we emphasize the O-passivated GNM system as it has not been reported in the literature. LDA predicts a gap of 0.7 eV for H-GNM, and 0.4 eV for O-GNM, suggesting that the passivation has some effect on the electronic structure of GNM's especially those with small pores. The bands of the passivated GNM's are linear away from the gap region, although with a maximal group velocity that is half that of graphene. The DOS of H-GNM, O-GNM and graphene are shown in Fig.~\ref{figure2}c. Projections of the DOS (PDOS) on the atomic orbitals shows that neither passivating species have a significant contribution at the band edges or in the linear region, and therefore one does not expect them to cause any resonant scattering.\cite{geimresonantscattering} Therefore GNM's, if doped up to the linear portion of the band structure, may have mobilities comparable to those based on pristine graphene, which can be exploited in novel devices. 

\begin{figure}
   \includegraphics[width=3.5in]{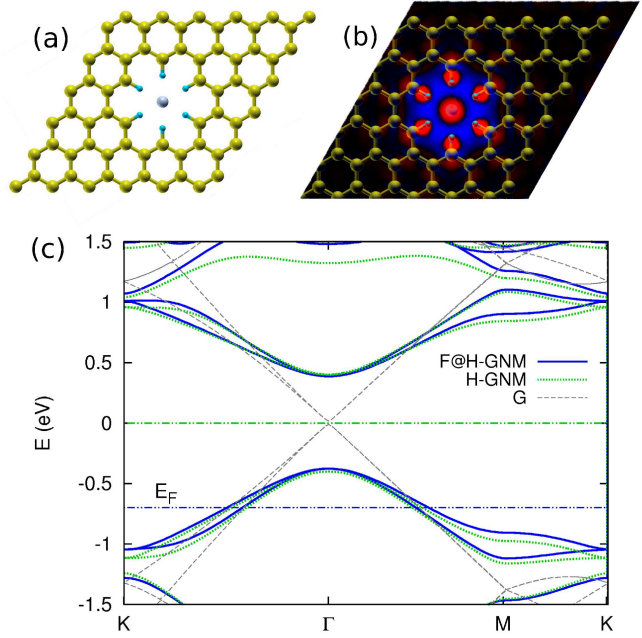}
   \caption
{ 
 {\it p}-doped GNM system. {\bf (a)} Unit cell of a hydrogen passivated GNM with fluorine chelation(F@H-GNM), constructed from a 6$\times$6 supercell of graphene, with a pore size of about 0.6 nm, and one fluorine atom per supercell. {\bf (b)} A 2D map of the charge density {\it difference} between the doped passivated system and its neutral components (GNM, hydrogen, and fluorine). The atomic positions are overlayed on the map. Red (blue) color indicates an increase (decrease) of electronic charge density. The red center shows that the fluorine is ionized. The blue color surrounding the center indicates that charge density is repelled away towards the pore-perimeter (due to the ionization of the fluorine atom). {\bf (c)} Band structures of: the doped F@H-GNM system (blue, solid); with its Fermi energy in the valence band (blue, dashed-dotted line), the undoped O-GNM system (green, dotted); with its Fermi energy at the origin (green, dashed-dotted line), and pristine graphene(grey, dashed). 
}
\label{figure3}
\end{figure}

\subsection{{\it p}-Doping}
We now turn to chelation doping as previously described (see Fig.~\ref{figure1}). Figure~\ref{figure3}a shows our first {\it p}-doped system. If a fluorine atom is brought close to the pore of a H-GNM (Fig.~\ref{figure3}a), we find that the system can lower its total energy if the fluorine docks in the center of the pore, forming a F@H-GNM, whilst an electron is transferred from the graphene sheet to the fluorine, ionizing it, and doping the GNM. The resulting fluorine {\it ion} is now electrostatically bound to the pore (there is some small concomitant lattice relaxation).

In support of our intuitive picture of the thermodynamics of docking (Fig.~\ref{figure1}), we show in Fig.~\ref{figure3}b a 2D map of the charge density {\it difference}, $\delta \sigma$, between the chelated/doped system and its neutral/undoped components (F and H-GNM),
\begin{equation}
\delta \sigma = \sigma_{\mathrm{F@H \mbox{-} GNM}} - \sigma_{\mathrm{H \mbox{-} GNM}} - \sigma_{\mathrm{F}}, 
\end{equation}
where $\sigma$ is the 2D charge density obtained by integrating the volume charge density over the direction perpendicular to the plane of the GNM. The 2D map shows that the fluorine is ionized. The extra electron on the fluorine polarizes the C-H bond leading to an induced dipole moment (blue-red regions around the pore in Fig.~\ref{figure3}b); the transferred carrier is {\it not} localized at the pore perimeter, rather, from L\"owdin analysis it can be ascertained that the donated carrier is delocalized across the GNM skeleton. 

To estimate the chemical stability of X@Y-GNM, we calculate the (zero-temperature) binding energy of the chelated crown GNM's using:
\begin{equation}
{E}_{\mathrm{binding}} = E_{\mathrm{X@Y \mbox{-} GNM}} - E_{\mathrm{Y \mbox{-} GNM}} - E_{\mathrm{X}}
\end{equation}
where $E_{\mathrm{X@Y \mbox{-} GNM}}$ is the energy of the passivated GNM system with the chelated ion, $E_{\mathrm{Y \mbox{-} GNM}}$ is the energy of the passivated GNM, and $E_{\mathrm{X}}$ is that of the isolated atom. For the F@H-GNM system, LDA gives a binding energy of $4.2$ eV, indicating very high thermal stability. 

A comparison of the band structures of the doped and undoped systems (Fig.~\ref{figure3}c) shows that ion chelation doping occurs within the rigid-band model. There is no significant change in the band curvatures, and the two band structures are essentially identical in the region of interest. The Fermi level of the chelated system is in the linear region of the valence band, 0.7 eV away from the middle of the gap.  Therefore, chelation doping of the crown GNM occurs such that the main features of the undoped system's electronic structure are preserved. 

\begin{figure}
\includegraphics[width=3.5in]{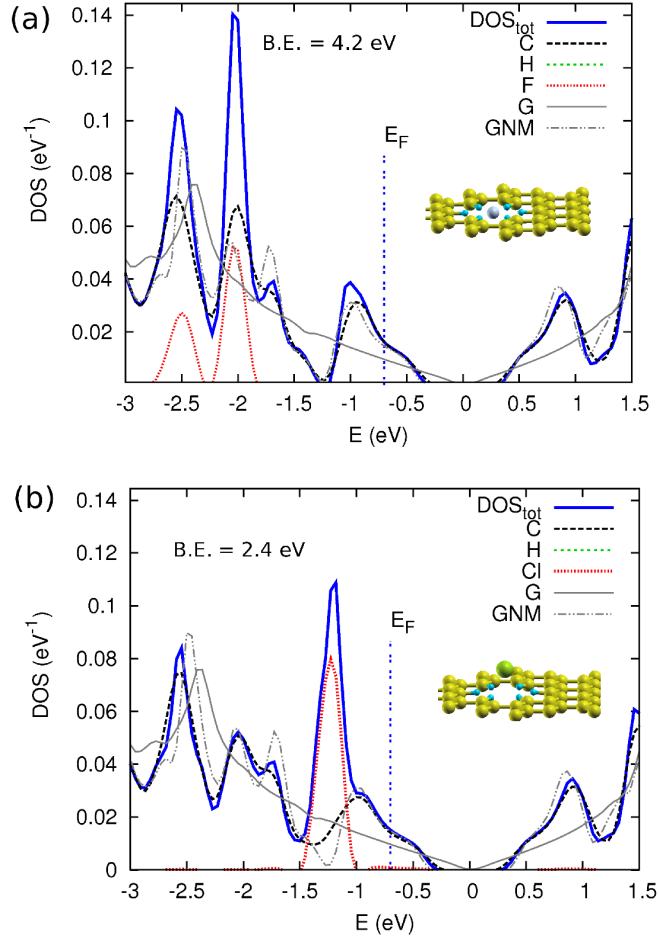}
\caption{ 
{\bf (a)} Total and projected DOS of the {\it p}-doped F@H-GNM system. For the total DOS (blue, solid), the Fermi level is below the gap (vertical line, blue dotted), indicating the {\it p}-doping of the GNM. Various contributions to the total DOS are shown; carbon 2{\it p}$_z$ states (black, dashed), hydrogen 1{\it s} states (green, small-dashed), and fluorine 2{\it p} states (red, dotted). Hydrogen has a negligible contribution in the energy range shown. The first fluorine peak, located approximately 1.25 eV below the Fermi level, corresponds to 2{\it p}$_x$ and 2{\it p}$_y$ states, while the one lower in energy is mainly a 2{\it p}$_z$ state. For comparison, undoped H-GNM total DOS is shown (grey, dashed-dotted), as well as pristine graphene (grey, solid). {\bf (b)} Total and projected DOS of the {\it p}-doped Cl@H-GNM system. Same notation as in {\bf (a)} with the F $\rightarrow$ Cl. The chlorine peak located 0.5 eV below the Fermi level corresponds to its 3{\it p} states.
} 
   \label{figure4}
\end{figure}

The rigid-band doping picture is confirmed by inspecting the DOS of the F@H-GNM system. Figure~\ref{figure4}a shows the total DOS and PDOS of that system,  with contributions from carbon, hydrogen, and fluorine states. The total DOS of the H-GNM and that of pristine graphene are also shown for comparison. The Fermi level of the F@H-GNM system indicates that the GNM is now {\it p}-doped. The DOS of the chelated and unchelated systems are very similar, except for the two peaks on the left side of the gap, which correspond to the fluorine {\it occupied} 2{\it p} states. These are located about 1.6 eV below the valence band edge, and about 1.3 eV from the Fermi level, and are therefore too far to interfere with low energy transport.

We next discuss our results for the second {\it p}-doping case; the chlorine chelated system, Cl@H-GNM. Here, the anionic radius of chlorine is too large for it to fit in the GNM, and the nanopore chelates the chlorine ion (radius=1.77\AA) at a distance 1.6 \AA \space above the plane of the GNM (Fig.~\ref{figure4}b, inset). Of course, a H-GNM with a larger pore radius could be constructed to comfortably accommodate chlorine. The electronic band structure of the Cl@H-GNM system is qualitatively similar to that of the F@H-GNM, with the Fermi energy in the valence band. Thus, chlorine chelation also occurs within the rigid-band doping model. Figure~\ref{figure4}b shows the total DOS and PDOS for the Cl@H-GNM system. The position of the Fermi energy indicates that the GNM is {\it p}-doped, without any significant chlorine contribution in the electronic states in the neighborhood of the Fermi energy, and up to the valence band edge. The chlorine 3{\it p} states are occupied. Close to the Fermi energy, we see a trace of chlorine states. These are found to be the 3{\it p}$_x$ and 3{\it p}$_y$ states, and their existence is a manifestation of the chlorine chelation above the plane of the pore (hence the slight appearance in the $\pi$ bands of the GNM). LDA predicts the binding energy of the Cl@H-GNM system to be 2.4 eV. This is smaller than that of fluorine due to steric effects pushing the chlorine out of plane. 

\subsection{{\it n}-doping}
We now turn to the {\it n}-doping of crown GNM's. This is achieved by a reversal of the {\it chemical logic} underlying the {\it p}-doping approach. Oxygen, which has an electronegativity higher than carbon, is used as the pore perimeter passivator. This leads to carbon-oxygen bond polarization, with more negative charge at the oxygen side, making the pore suitable for hosting positive ions. Two electron-donating elements, potassium and sodium, are used to demonstrate {\it n}-doping of the O-GNM system. Due to the ionic sizes of these two elements, a GNM system with a bigger unit cell is used to accommodate for a larger pore. 

\begin{figure}
   \includegraphics[width=3.5in]{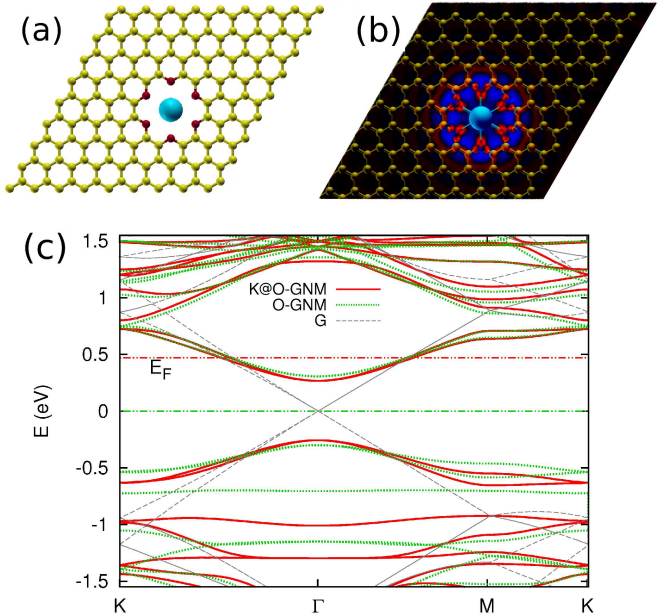}
   \caption
{ 
 {\it n}-doped GNM system. {\bf (a)}. Unit cell of an oxygen passivated GNM with potassium chelation(K@O-GNM), constructed from a 9$\times$9 supercell of graphene. {\bf (b)} A 2D map of the charge density {\it difference} between the doped passivated system and its neutral components (GNM, oxygen, and potassium). The atomic positions are overlayed on the map. The blue center shows that the potassium atom is ionized. The red color on the pore side of the oxygen atoms indicates that charge density is attracted towards the center of the pore (due to the ionization of the potassium atom). {\bf (c)}. Band structures of: the doped K@O-GNM system (red, solid); with its Fermi energy in the conduction band (red, dashed-dotted line), the undoped O-GNM system (green, dotted); with its Fermi energy at the origin (green, dashed-dotted line), and pristine graphene(grey, dashed). 
}
   \label{figure5}
\end{figure}

Figure~\ref{figure5}a, shows an oxygen-passivated GNM with a pore size of $0.7$ nm. The relaxed configuration of the unchelated (undoped) O-GNM has a band gap of $0.4$ eV. By bringing the potassium atom to a distance of a few angstroms from the pore, the system lowers its energy by the docking and the ionization of the potassium atom in the center of the pore, with its 4s electron donated to the O-GNM $\pi$ bands (see Fig.~\ref{figure1}). Figure~\ref{figure5}b shows the 2D charge density difference map for the system, as previously defined. The central region indicates that potassium atom has been ionized, with accompanying charge polarization induced at the pore perimeter. LDA predicts that the potassium is bound to the GNM by a binding energy of $2.8$ eV.
 
\begin{figure}
   \includegraphics[width=3.5in]{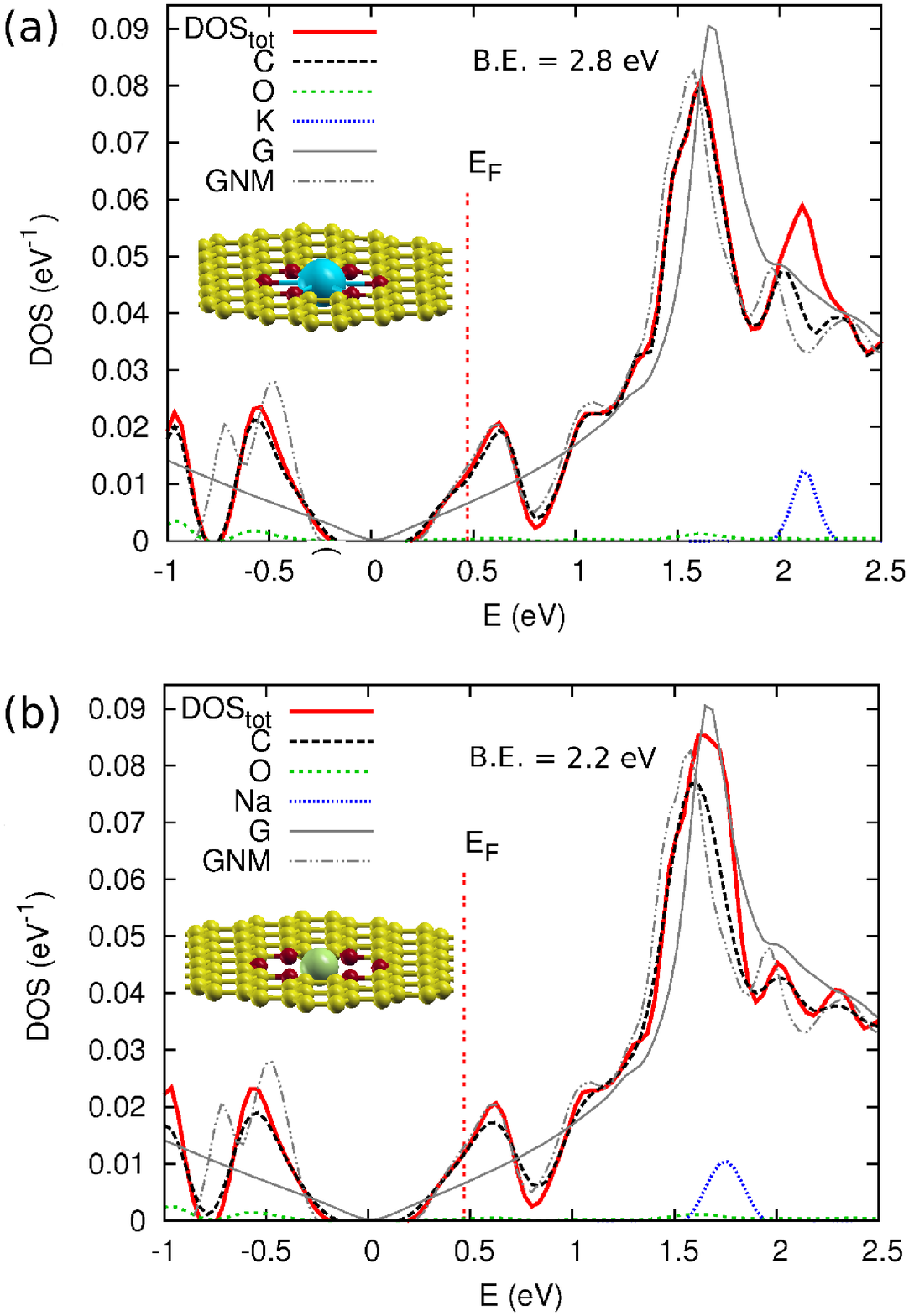}
   \caption
{ 
{\bf (a)} Total and projected DOS of the {\it n}-doped K@O-GNM system. For the total DOS (red, solid), the Fermi level is above the gap (vertical line, red dotted), indicating the {\it n}-doping of the GNM. Various contributions to the total DOS are shown; carbon 2{\it p}$_z$ states (black, dashed), oxygen 2{\it p} states (green, small-dashed), and potassium 4{\it s} states (blue, dotted). Oxygen has a negligible contribution in the energy range of interest. The potassium peak located approximately 1.5 eV above the Fermi level corresponds to its 4{\it s} state. For comparison, undoped O-GNM total DOS is shown (grey, dashed-dotted), as well as pristine graphene (grey, solid). {\bf (b)} Total and projected DOS of the {\it n}-doped Na@O-GNM system. Same notation as in {\bf (a)} with the K $\rightarrow$ Na. The sodium 3{\it s} state is located 1.25 eV above the Fermi level.
}
   \label{figure6}
\end{figure}

Comparing the band structure of K@O-GNM and O-GNM shows that potassium chelation preserves the O-GNM band structure(Fig.~\ref{figure5}c) in the region of interest. The chelation merely shifts the Fermi level into the conduction band. This observation is further confirmed by inspection of the DOS and PDOS of the K@O-GNM system (Fig.~\ref{figure6}a), where we see that the DOS's of the two systems are very similar. The potassium 4s level does not contribute to the spectrum in the vicinity of the gap.

We continue our examination of X@Y-GNM {\it n}-doping by performing a study of a sodium-chelated system. The minimum energy configuration of the Na@O-GNM system binds the Na ion in the center of the pore with an energy of 2.2 eV. The band structure of the Na@O-GNM system is almost identical to that of the K@O-GNM, also with the Fermi energy in the conduction band of the O-GNM, indicating {\it n}-doping. The total DOS and PDOS of the Na@O-GNM system (Fig.~\ref{figure6}b) confirms the same physics is operative here, as in the K@O-GNM case, with the empty sodium 3{\it s} state at about $1.25$eV above the Fermi level.

We observe high binding energies for the studied {\it p}- and {\it n}-doped crown GNM systems, a desirable feature for a material to be used in technological applications. It is therefore useful to describe this physics in more detail. First, consider the energetics of chelant ionization; the creation of negative ions generally results in a release of energy whilst the creation of positive ions costs energy. Second, consider the attractive electrostatic interaction between the X ion and the Y-GNM. Although fluorine has a lower electron affinity than chlorine, the F@H-GNM has a higher binding energy than the Cl@H-GNM due to the larger attractive electrostatic interaction of the fluorine ion with the hydrogen-lined pore. Chlorine's large ionic radius keeps it 1.67\AA~above the pore (exchange repulsion) preventing the full electrostatic gain. On the other hand, potassium has a lower ionization energy than sodium, and since they both dock at the pore center, they have approximately equal electrostatic interaction energies with the pore. As a result, K@O-GNM has a higher binding energy than Na@O-GNM by approximately the difference between their first ionization energies. More generally, these two factors will determine the binding energy, hence the stability, of a chelant ion in a passivated GNM, the lattice relaxation and the electron affinity/ionization energy of the GNM being small in comparison. It should also be noted here that chelated dopants are very stable compared to adsorbed ones, \cite{Tapia2011,Lugo-Solis2007} with a binding energy of about 0.1 eV for the adsorbed potassium atom atop a graphene surface.

The presented doping picture is not expected to vary qualititively with the density of the pores. Decreasing the number of pores in the graphene lattice (i.e. increasing the pore-pore distance) lowers the electronic cost for hosting the transfered charge, as there are more carbon atoms which can share the charge transfered from the chelant atom. Unless one has a high density of pores (where most of the graphene lattice is removed), the implications on the binding energy will be small. Systems addressing this point are now being studied.

More advanced calculations using hybrid density functionals such as B3LYP\cite{frisch,LeeParr,Becke1993}, PBE0\cite{Adamo1999}, and HSE\cite{HSE} have proved to be more quantitatively accurate for the calculation of the electronic properties of molecular systems. In particular, the HSE functional has been successful in determining the electronic properties of carbon nanotubes and graphene nanoribbon systems.\cite{Barone2011} We performed calculations using the HSE hybrid functional for the F@H-GNM system, and the results confirm the message conveyed by LDA and GGA (see supplementary information).

The X@Y-GNM chelation doping approach offers a high degree of concentration control at the nanoscale. The studied cases assume one chelant per unit cell of the GNM lattice. As shown, GNM doping occurs within a rigid-band picture, and therefore an {\it ordered} partial-loading of the pores would introduce fewer carriers into the graphene lattice. The strong binding between the dopant (chelant) and the pore guarantees the structural stability.

\section{CONCLUSION} 

Graphene nanomeshes (GNM's), due to their acceptably large band gaps and high mobilities, have great potential for use in nanoelectronics applications. Here we have shown using first-principles calculations that the chemically motivated ion chelation approach is suitable for achieving stable controlled doping in GNM's. This doping mechanism is based on the selective passivation of GNM's pore perimeters which allows for the preferential "docking" of positive or negative ions and hence specific doping. Furthermore, unlike other approaches, chelation doping offers both ultra stability ($> 80 k_BT$ at room temperature) and a precise control over carrier type and concentration. We also demonstrate that ion chelation preserves the electronic structure of GNM's by merely moving the Fermi level to their valence or conduction bands, the rigid-band mechanism. Realization of chelated GNM's would enable a wide-spread set of important nanoelectronic applications.
 
\section{METHODS} 

Band structure calculations were performed with the Quantum Espresso software package\textregistered on relaxed configurations (with forces less than 0.001 Ry/Bohr). The electronic structure calculations were performed on up to 512 nodes of IBM's T. J. Watson BlueGene/P supercomputer using the Purdew-Zunger local density approximation functional (PZ-LDA), \cite{PZLDA} with an energy cutoff of 45 Ry, a $12 \times 12 \times 1$ Monkhorst-Pack k-point grid for the X@H-GNM systems, and a  $6 \times 6 \times 1$ k-point grid for the larger X@O-GNM ones. Images are separated along the $z$-direction by 11 \AA.

Pseudopotentials for H, C, K, and Na were of the normconserving Von-Barth-Car direct-fit type, Cl was of the norm conserving Bachelet-Hamann-Schl\"uter type, and O and F were of the ultrasoft type. The charge transferred from/to the GNM systems studied is calculated by integrating the total DOS of the system. 

\acknowledgement
The authors thank Ph. Avouris, G. Tulevski, A. Aboukandil, and H. Mohamed for useful discussions. This work was supported by funding from STDF and ITIDA (Egypt), as well as by the computational resources at IBM T. J. Watson Research Center.

\suppinfo

Results for calculations using other flavors of DFT are also presented for exemplar systems. The electronic properties of unpassivated GNM's are also discussed.

\bibliography{GNMPapersAll}

\providecommand*\mcitethebibliography{\thebibliography}
\csname @ifundefined\endcsname{endmcitethebibliography}
  {\let\endmcitethebibliography\endthebibliography}{}
\begin{mcitethebibliography}{47}
\providecommand*\natexlab[1]{#1}
\providecommand*\mciteSetBstSublistMode[1]{}
\providecommand*\mciteSetBstMaxWidthForm[2]{}
\providecommand*\mciteBstWouldAddEndPuncttrue
  {\def\EndOfBibitem{\unskip.}}
\providecommand*\mciteBstWouldAddEndPunctfalse
  {\let\EndOfBibitem\relax}
\providecommand*\mciteSetBstMidEndSepPunct[3]{}
\providecommand*\mciteSetBstSublistLabelBeginEnd[3]{}
\providecommand*\EndOfBibitem{}
\mciteSetBstSublistMode{f}
\mciteSetBstMaxWidthForm{subitem}{(\alph{mcitesubitemcount})}
\mciteSetBstSublistLabelBeginEnd
  {\mcitemaxwidthsubitemform\space}
  {\relax}
  {\relax}

\bibitem[Castro~Neto et~al.(2009)Castro~Neto, Guinea, Peres, Novoselov, and
  Geim]{neto2009}
Castro~Neto,~A.~H.; Guinea,~F.; Peres,~N. M.~R.; Novoselov,~K.~S.; Geim,~A.~K.
  \emph{Rev. Mod. Phys.} \textbf{2009}, \emph{81}, 109--162\relax
\mciteBstWouldAddEndPuncttrue
\mciteSetBstMidEndSepPunct{\mcitedefaultmidpunct}
{\mcitedefaultendpunct}{\mcitedefaultseppunct}\relax
\EndOfBibitem
\bibitem[Geim and Novoselov(2007)Geim, and Novoselov]{geim2007}
Geim,~A.~K.; Novoselov,~K.~S. \emph{Nat. Mater.} \textbf{2007}, \emph{6},
  183--191\relax
\mciteBstWouldAddEndPuncttrue
\mciteSetBstMidEndSepPunct{\mcitedefaultmidpunct}
{\mcitedefaultendpunct}{\mcitedefaultseppunct}\relax
\EndOfBibitem
\bibitem[Sire et~al.(2012)Sire, Ardiaca, Lepilliet, Seo, Hersam, Dambrine,
  Happy, and Derycke]{graphenetrans1}
Sire,~C.; Ardiaca,~F.; Lepilliet,~S.; Seo,~J.-W.~T.; Hersam,~M.~C.;
  Dambrine,~G.; Happy,~H.; Derycke,~V. \emph{Nano Lett.} \textbf{2012},
  \emph{12}, 1184--8\relax
\mciteBstWouldAddEndPuncttrue
\mciteSetBstMidEndSepPunct{\mcitedefaultmidpunct}
{\mcitedefaultendpunct}{\mcitedefaultseppunct}\relax
\EndOfBibitem
\bibitem[Lin et~al.(2009)Lin, Jenkins, Valdes-Garcia, Small, Farmer, and
  Avouris]{graphenetrans2}
Lin,~Y.-M.; Jenkins,~K.~A.; Valdes-Garcia,~A.; Small,~J.~P.; Farmer,~D.~B.;
  Avouris,~P. \emph{Nano Lett.} \textbf{2009}, \emph{9}, 422--426\relax
\mciteBstWouldAddEndPuncttrue
\mciteSetBstMidEndSepPunct{\mcitedefaultmidpunct}
{\mcitedefaultendpunct}{\mcitedefaultseppunct}\relax
\EndOfBibitem
\bibitem[Lin et~al.(2010)Lin, Dimitrakopoulos, Jenkins, Farmer, Chiu, Grill,
  and Avouris]{graphenetrans3}
Lin,~Y.~M.; Dimitrakopoulos,~C.; Jenkins,~K.~A.; Farmer,~D.~B.; Chiu,~H.~Y.;
  Grill,~A.; Avouris,~P. \emph{Science} \textbf{2010}, \emph{327}, 662\relax
\mciteBstWouldAddEndPuncttrue
\mciteSetBstMidEndSepPunct{\mcitedefaultmidpunct}
{\mcitedefaultendpunct}{\mcitedefaultseppunct}\relax
\EndOfBibitem
\bibitem[Wu et~al.(2011)Wu, Lin, Bol, Jenkins, Xia, Farmer, Zhu, and
  Avouris]{graphenetrans4}
Wu,~Y.; Lin,~Y.-m.; Bol,~A.~a.; Jenkins,~K.~a.; Xia,~F.; Farmer,~D.~B.;
  Zhu,~Y.; Avouris,~P. \emph{Nature} \textbf{2011}, \emph{472}, 74--8\relax
\mciteBstWouldAddEndPuncttrue
\mciteSetBstMidEndSepPunct{\mcitedefaultmidpunct}
{\mcitedefaultendpunct}{\mcitedefaultseppunct}\relax
\EndOfBibitem
\bibitem[Jiang et~al.(2009)Jiang, Cooper, and Dai]{Jiang2009}
Jiang,~D.-e.; Cooper,~V.~R.; Dai,~S. \emph{Nano Lett.} \textbf{2009}, \emph{9},
  4019--24\relax
\mciteBstWouldAddEndPuncttrue
\mciteSetBstMidEndSepPunct{\mcitedefaultmidpunct}
{\mcitedefaultendpunct}{\mcitedefaultseppunct}\relax
\EndOfBibitem
\bibitem[Machado and Serp(2012)Machado, and Serp]{graphenecatalysis}
Machado,~B.~F.; Serp,~P. \emph{Catal. Sci. Technol.} \textbf{2012}, \emph{2},
  54--75\relax
\mciteBstWouldAddEndPuncttrue
\mciteSetBstMidEndSepPunct{\mcitedefaultmidpunct}
{\mcitedefaultendpunct}{\mcitedefaultseppunct}\relax
\EndOfBibitem
\bibitem[Xia et~al.(2010)Xia, Farmer, Lin, and Avouris]{Xia2010}
Xia,~F.; Farmer,~D.~B.; Lin,~Y.-M.; Avouris,~P. \emph{Nano Lett.}
  \textbf{2010}, \emph{10}, 715--8\relax
\mciteBstWouldAddEndPuncttrue
\mciteSetBstMidEndSepPunct{\mcitedefaultmidpunct}
{\mcitedefaultendpunct}{\mcitedefaultseppunct}\relax
\EndOfBibitem
\bibitem[Malola et~al.(2009)Malola, Häkkinen, and Koskinen]{Malola2009}
Malola,~S.; Häkkinen,~H.; Koskinen,~P. \emph{Appl. Phys. Lett.} \textbf{2009},
  \emph{94}, 043106\relax
\mciteBstWouldAddEndPuncttrue
\mciteSetBstMidEndSepPunct{\mcitedefaultmidpunct}
{\mcitedefaultendpunct}{\mcitedefaultseppunct}\relax
\EndOfBibitem
\bibitem[Sanyal et~al.(2009)Sanyal, Eriksson, Jansson, and
  Grennberg]{Sanyal2009}
Sanyal,~B.; Eriksson,~O.; Jansson,~U.; Grennberg,~H. \emph{Phys. Rev. B}
  \textbf{2009}, \emph{79}, 113409\relax
\mciteBstWouldAddEndPuncttrue
\mciteSetBstMidEndSepPunct{\mcitedefaultmidpunct}
{\mcitedefaultendpunct}{\mcitedefaultseppunct}\relax
\EndOfBibitem
\bibitem[Topsakal et~al.(2008)Topsakal, Akt\"{u}rk, Sevin\c{c}li, and
  Ciraci]{Topsakal2008}
Topsakal,~M.; Akt\"{u}rk,~E.; Sevin\c{c}li,~H.; Ciraci,~S. \emph{Phys. Rev. B}
  \textbf{2008}, \emph{78}, 235435\relax
\mciteBstWouldAddEndPuncttrue
\mciteSetBstMidEndSepPunct{\mcitedefaultmidpunct}
{\mcitedefaultendpunct}{\mcitedefaultseppunct}\relax
\EndOfBibitem
\bibitem[Pisani et~al.(2008)Pisani, Montanari, and Harrison]{Pisani2008}
Pisani,~L.; Montanari,~B.; Harrison,~N.~M. \emph{New J. Phys.} \textbf{2008},
  \emph{10}, 033002\relax
\mciteBstWouldAddEndPuncttrue
\mciteSetBstMidEndSepPunct{\mcitedefaultmidpunct}
{\mcitedefaultendpunct}{\mcitedefaultseppunct}\relax
\EndOfBibitem
\bibitem[F\"{u}rst et~al.(2009)F\"{u}rst, Pedersen, Flindt, Mortensen,
  Brandbyge, Pedersen, and a~P~Jauho]{Furst2009}
F\"{u}rst,~J.~a.; Pedersen,~J.~G.; Flindt,~C.; Mortensen,~N.~a.; Brandbyge,~M.;
  Pedersen,~T.~G.; a~P~Jauho, \emph{New J. Phys.} \textbf{2009}, \emph{11},
  095020\relax
\mciteBstWouldAddEndPuncttrue
\mciteSetBstMidEndSepPunct{\mcitedefaultmidpunct}
{\mcitedefaultendpunct}{\mcitedefaultseppunct}\relax
\EndOfBibitem
\bibitem[Pedersen et~al.(2008)Pedersen, Flindt, Pedersen, Mortensen, Jauho, and
  Pedersen]{Pedersen2008}
Pedersen,~T.; Flindt,~C.; Pedersen,~J.; Mortensen,~N.; Jauho,~A.-P.;
  Pedersen,~K. \emph{Phys. Rev. Lett.} \textbf{2008}, \emph{100}, 1--4\relax
\mciteBstWouldAddEndPuncttrue
\mciteSetBstMidEndSepPunct{\mcitedefaultmidpunct}
{\mcitedefaultendpunct}{\mcitedefaultseppunct}\relax
\EndOfBibitem
\bibitem[Liu et~al.(2009)Liu, Wang, Shi, Yang, and Liu]{Liu2009}
Liu,~W.; Wang,~Z.~F.; Shi,~Q.~W.; Yang,~J.; Liu,~F. \emph{Phys. Rev. B}
  \textbf{2009}, \emph{80}, 2--5\relax
\mciteBstWouldAddEndPuncttrue
\mciteSetBstMidEndSepPunct{\mcitedefaultmidpunct}
{\mcitedefaultendpunct}{\mcitedefaultseppunct}\relax
\EndOfBibitem
\bibitem[Petersen and Pedersen(2009)Petersen, and Pedersen]{Petersen2009}
Petersen,~R.; Pedersen,~T. \emph{Phys. Rev. B} \textbf{2009}, \emph{80},
  1--4\relax
\mciteBstWouldAddEndPuncttrue
\mciteSetBstMidEndSepPunct{\mcitedefaultmidpunct}
{\mcitedefaultendpunct}{\mcitedefaultseppunct}\relax
\EndOfBibitem
\bibitem[Vanevi\'{c} et~al.(2009)Vanevi\'{c}, Stojanovi\'{c}, and
  Kindermann]{Vanevic2009}
Vanevi\'{c},~M.; Stojanovi\'{c},~V.; Kindermann,~M. \emph{Phys. Rev. B}
  \textbf{2009}, \emph{80}, 1--8\relax
\mciteBstWouldAddEndPuncttrue
\mciteSetBstMidEndSepPunct{\mcitedefaultmidpunct}
{\mcitedefaultendpunct}{\mcitedefaultseppunct}\relax
\EndOfBibitem
\bibitem[Martinazzo et~al.(2010)Martinazzo, Casolo, and
  Tantardini]{Martinazzo2010}
Martinazzo,~R.; Casolo,~S.; Tantardini,~G. \emph{Phys. Rev. B} \textbf{2010},
  \emph{81}, 1--8\relax
\mciteBstWouldAddEndPuncttrue
\mciteSetBstMidEndSepPunct{\mcitedefaultmidpunct}
{\mcitedefaultendpunct}{\mcitedefaultseppunct}\relax
\EndOfBibitem
\bibitem[Baskin and Kr\'{a}l(2011)Baskin, and Kr\'{a}l]{Baskin2011}
Baskin,~A.; Kr\'{a}l,~P. \emph{Sci. Rep.} \textbf{2011}, \emph{1}, 36\relax
\mciteBstWouldAddEndPuncttrue
\mciteSetBstMidEndSepPunct{\mcitedefaultmidpunct}
{\mcitedefaultendpunct}{\mcitedefaultseppunct}\relax
\EndOfBibitem
\bibitem[Bai et~al.(2010)Bai, Zhong, Jiang, Huang, and Duan]{Bai2010}
Bai,~J.; Zhong,~X.; Jiang,~S.; Huang,~Y.; Duan,~X. \emph{Nat. Nanotechnol.}
  \textbf{2010}, \emph{5}, 190--4\relax
\mciteBstWouldAddEndPuncttrue
\mciteSetBstMidEndSepPunct{\mcitedefaultmidpunct}
{\mcitedefaultendpunct}{\mcitedefaultseppunct}\relax
\EndOfBibitem
\bibitem[Safron et~al.(2011)Safron, Brewer, and Arnold]{Safron2011}
Safron,~N.~S.; Brewer,~A.~S.; Arnold,~M.~S. \emph{Small} \textbf{2011},
  \emph{7}, 492--8\relax
\mciteBstWouldAddEndPuncttrue
\mciteSetBstMidEndSepPunct{\mcitedefaultmidpunct}
{\mcitedefaultendpunct}{\mcitedefaultseppunct}\relax
\EndOfBibitem
\bibitem[Liu et~al.(2011)Liu, Zhang, Wang, Gu, Bai, and Wang]{Liu2011}
Liu,~L.; Zhang,~Y.; Wang,~W.; Gu,~C.; Bai,~X.; Wang,~E. \emph{Adv. Mater.}
  \textbf{2011}, \emph{23}, 1246--51\relax
\mciteBstWouldAddEndPuncttrue
\mciteSetBstMidEndSepPunct{\mcitedefaultmidpunct}
{\mcitedefaultendpunct}{\mcitedefaultseppunct}\relax
\EndOfBibitem
\bibitem[Liang et~al.(2010)Liang, Jung, Wu, Ismach, Olynick, Cabrini, and
  Bokor]{Liang2010}
Liang,~X.; Jung,~Y.-S.; Wu,~S.; Ismach,~A.; Olynick,~D.~L.; Cabrini,~S.;
  Bokor,~J. \emph{Nano Lett.} \textbf{2010}, \emph{10}, 2454--60\relax
\mciteBstWouldAddEndPuncttrue
\mciteSetBstMidEndSepPunct{\mcitedefaultmidpunct}
{\mcitedefaultendpunct}{\mcitedefaultseppunct}\relax
\EndOfBibitem
\bibitem[Liu et~al.(2011)Liu, Liu, and Zhu]{graphenedoping1}
Liu,~H.; Liu,~Y.; Zhu,~D. \emph{J. Mater. Chem.} \textbf{2011}, \emph{21},
  3335\relax
\mciteBstWouldAddEndPuncttrue
\mciteSetBstMidEndSepPunct{\mcitedefaultmidpunct}
{\mcitedefaultendpunct}{\mcitedefaultseppunct}\relax
\EndOfBibitem
\bibitem[Qu et~al.(2010)Qu, Liu, Baek, and Dai]{graphenenitrogendoping1}
Qu,~L.; Liu,~Y.; Baek,~J.-B.; Dai,~L. \emph{ACS nano} \textbf{2010}, \emph{4},
  1321--6\relax
\mciteBstWouldAddEndPuncttrue
\mciteSetBstMidEndSepPunct{\mcitedefaultmidpunct}
{\mcitedefaultendpunct}{\mcitedefaultseppunct}\relax
\EndOfBibitem
\bibitem[Wang et~al.(2009)Wang, Li, Zhang, Yoon, Weber, Wang, Guo, and
  Dai]{graphenenitrogendoping2}
Wang,~X.; Li,~X.; Zhang,~L.; Yoon,~Y.; Weber,~P.~K.; Wang,~H.; Guo,~J.; Dai,~H.
  \emph{Science} \textbf{2009}, \emph{324}, 768--71\relax
\mciteBstWouldAddEndPuncttrue
\mciteSetBstMidEndSepPunct{\mcitedefaultmidpunct}
{\mcitedefaultendpunct}{\mcitedefaultseppunct}\relax
\EndOfBibitem
\bibitem[Firlej et~al.(2009)Firlej, Kuchta, Wexler, and
  Pfeifer]{grapheneborondoping1}
Firlej,~L.; Kuchta,~B.; Wexler,~C.; Pfeifer,~P. \emph{Adsorption}
  \textbf{2009}, \emph{15}, 312--317\relax
\mciteBstWouldAddEndPuncttrue
\mciteSetBstMidEndSepPunct{\mcitedefaultmidpunct}
{\mcitedefaultendpunct}{\mcitedefaultseppunct}\relax
\EndOfBibitem
\bibitem[Kim et~al.(2010)Kim, Reina, Shi, Park, Li, Lee, and Kong]{kim2010}
Kim,~K.~K.; Reina,~A.; Shi,~Y.; Park,~H.; Li,~L.-J.; Lee,~Y.~H.; Kong,~J.
  \emph{Nanotechnology} \textbf{2010}, \emph{21}, 285205\relax
\mciteBstWouldAddEndPuncttrue
\mciteSetBstMidEndSepPunct{\mcitedefaultmidpunct}
{\mcitedefaultendpunct}{\mcitedefaultseppunct}\relax
\EndOfBibitem
\bibitem[Hua and Flood(2010)Hua, and Flood]{hua2010}
Hua,~Y.; Flood,~A.~H. \emph{Chem. Soc. Rev.} \textbf{2010}, \emph{39},
  1262--1271\relax
\mciteBstWouldAddEndPuncttrue
\mciteSetBstMidEndSepPunct{\mcitedefaultmidpunct}
{\mcitedefaultendpunct}{\mcitedefaultseppunct}\relax
\EndOfBibitem
\bibitem[Li and Flood(2008)Li, and Flood]{indianamolecule}
Li,~Y.; Flood,~A. \emph{Angew. Chem., Int. Ed.} \textbf{2008}, \emph{47},
  2649--2652\relax
\mciteBstWouldAddEndPuncttrue
\mciteSetBstMidEndSepPunct{\mcitedefaultmidpunct}
{\mcitedefaultendpunct}{\mcitedefaultseppunct}\relax
\EndOfBibitem
\bibitem[Pedersen(1967)]{crownethers1}
Pedersen,~C.~J. \emph{J. Am. Chem. Soc.} \textbf{1967}, \emph{89},
  7017--7036\relax
\mciteBstWouldAddEndPuncttrue
\mciteSetBstMidEndSepPunct{\mcitedefaultmidpunct}
{\mcitedefaultendpunct}{\mcitedefaultseppunct}\relax
\EndOfBibitem
\bibitem[Christensen et~al.(1971)Christensen, Hill, and Izatt]{crownethers2}
Christensen,~J.~J.; Hill,~J.~O.; Izatt,~R.~M. \emph{Science} \textbf{1971},
  \emph{174}, 459--467\relax
\mciteBstWouldAddEndPuncttrue
\mciteSetBstMidEndSepPunct{\mcitedefaultmidpunct}
{\mcitedefaultendpunct}{\mcitedefaultseppunct}\relax
\EndOfBibitem
\bibitem[Nistor et~al.(2011)Nistor, Newns, and Martyna]{Nistor2011}
Nistor,~R.~A.; Newns,~D.~M.; Martyna,~G.~J. \emph{ACS nano} \textbf{2011},
  \emph{5}, 3096--3103\relax
\mciteBstWouldAddEndPuncttrue
\mciteSetBstMidEndSepPunct{\mcitedefaultmidpunct}
{\mcitedefaultendpunct}{\mcitedefaultseppunct}\relax
\EndOfBibitem
\bibitem[NEWNS(1969)]{newns1}
NEWNS,~D.~M. \emph{Phys. Rev.} \textbf{1969}, \emph{178}, 1123--1135\relax
\mciteBstWouldAddEndPuncttrue
\mciteSetBstMidEndSepPunct{\mcitedefaultmidpunct}
{\mcitedefaultendpunct}{\mcitedefaultseppunct}\relax
\EndOfBibitem
\bibitem[Hewson and Newns(1974)Hewson, and Newns]{newns2}
Hewson,~A.~C.; Newns,~D.~M. \emph{Jpn. J. Appl. Phys.} \textbf{1974},
  \emph{2S2}, 121--131\relax
\mciteBstWouldAddEndPuncttrue
\mciteSetBstMidEndSepPunct{\mcitedefaultmidpunct}
{\mcitedefaultendpunct}{\mcitedefaultseppunct}\relax
\EndOfBibitem
\bibitem[Wehling et~al.(2010)Wehling, Yuan, Lichtenstein, Geim, and
  Katsnelson]{geimresonantscattering}
Wehling,~T.~O.; Yuan,~S.; Lichtenstein,~A.~I.; Geim,~A.~K.; Katsnelson,~M.~I.
  \emph{Phys. Rev. Lett.} \textbf{2010}, \emph{105}, 056802\relax
\mciteBstWouldAddEndPuncttrue
\mciteSetBstMidEndSepPunct{\mcitedefaultmidpunct}
{\mcitedefaultendpunct}{\mcitedefaultseppunct}\relax
\EndOfBibitem
\bibitem[Tapia et~al.(2011)Tapia, Acosta, Medina-Esquivel, and
  Canto]{Tapia2011}
Tapia,~a.; Acosta,~C.; Medina-Esquivel,~R.; Canto,~G. \emph{Comput. Mater.
  Sci.} \textbf{2011}, \emph{50}, 2427--2432\relax
\mciteBstWouldAddEndPuncttrue
\mciteSetBstMidEndSepPunct{\mcitedefaultmidpunct}
{\mcitedefaultendpunct}{\mcitedefaultseppunct}\relax
\EndOfBibitem
\bibitem[Lugo-Solis and Vasiliev(2007)Lugo-Solis, and Vasiliev]{Lugo-Solis2007}
Lugo-Solis,~A.; Vasiliev,~I. \emph{Phys. Rev. B} \textbf{2007}, \emph{76},
  235431\relax
\mciteBstWouldAddEndPuncttrue
\mciteSetBstMidEndSepPunct{\mcitedefaultmidpunct}
{\mcitedefaultendpunct}{\mcitedefaultseppunct}\relax
\EndOfBibitem
\bibitem[Stephens et~al.(1994)Stephens, Devlin, Chabalowski, and
  Frisch]{frisch}
Stephens,~P.~J.; Devlin,~F.~J.; Chabalowski,~C.~F.; Frisch,~M.~J. \emph{J.
  Phys. Chem.} \textbf{1994}, \emph{98}, 11623--11627\relax
\mciteBstWouldAddEndPuncttrue
\mciteSetBstMidEndSepPunct{\mcitedefaultmidpunct}
{\mcitedefaultendpunct}{\mcitedefaultseppunct}\relax
\EndOfBibitem
\bibitem[Lee et~al.(1988)Lee, Yang, and Parr]{LeeParr}
Lee,~C.; Yang,~W.; Parr,~R.~G. \emph{Phys. Rev. B} \textbf{1988}, \emph{37},
  785--789\relax
\mciteBstWouldAddEndPuncttrue
\mciteSetBstMidEndSepPunct{\mcitedefaultmidpunct}
{\mcitedefaultendpunct}{\mcitedefaultseppunct}\relax
\EndOfBibitem
\bibitem[Becke(1993)]{Becke1993}
Becke,~A.~D. \emph{J. Chem. Phys.} \textbf{1993}, \emph{98}, 5648--5652\relax
\mciteBstWouldAddEndPuncttrue
\mciteSetBstMidEndSepPunct{\mcitedefaultmidpunct}
{\mcitedefaultendpunct}{\mcitedefaultseppunct}\relax
\EndOfBibitem
\bibitem[Adamo and Barone(1999)Adamo, and Barone]{Adamo1999}
Adamo,~C.; Barone,~V. \emph{J. Chem. Phys.} \textbf{1999}, \emph{110},
  6158\relax
\mciteBstWouldAddEndPuncttrue
\mciteSetBstMidEndSepPunct{\mcitedefaultmidpunct}
{\mcitedefaultendpunct}{\mcitedefaultseppunct}\relax
\EndOfBibitem
\bibitem[Heyd et~al.(2003)Heyd, Scuseria, and Ernzerhof]{HSE}
Heyd,~J.; Scuseria,~G.~E.; Ernzerhof,~M. \emph{J. Chem. Phys.} \textbf{2003},
  \emph{118}, 8207--8215\relax
\mciteBstWouldAddEndPuncttrue
\mciteSetBstMidEndSepPunct{\mcitedefaultmidpunct}
{\mcitedefaultendpunct}{\mcitedefaultseppunct}\relax
\EndOfBibitem
\bibitem[Barone et~al.(2011)Barone, Hod, Peralta, and Scuseria]{Barone2011}
Barone,~V.; Hod,~O.; Peralta,~J.~E.; Scuseria,~G.~E. \emph{Acc. Chem. Res.}
  \textbf{2011}, \emph{44}, 269--79\relax
\mciteBstWouldAddEndPuncttrue
\mciteSetBstMidEndSepPunct{\mcitedefaultmidpunct}
{\mcitedefaultendpunct}{\mcitedefaultseppunct}\relax
\EndOfBibitem
\bibitem[Perdew and Zunger(1981)Perdew, and Zunger]{PZLDA}
Perdew,~J.~P.; Zunger,~A. \emph{Phys. Rev. B} \textbf{1981}, \emph{23},
  5048--5079\relax
\mciteBstWouldAddEndPuncttrue
\mciteSetBstMidEndSepPunct{\mcitedefaultmidpunct}
{\mcitedefaultendpunct}{\mcitedefaultseppunct}\relax
\EndOfBibitem
\end{mcitethebibliography}

\end{document}